\documentclass[10pt]{article}
\usepackage{epsfig}
\begin{document}
\title{Transmission Properties of the oscillating delta-function
potential}

\author{D.F. Martinez and L.E. Reichl\\Center for Statistical Mechanics\\Department of Physics\\The
University of Texas at Austin\\Austin, TX, 78701,U.S.A\\}

\maketitle
\begin{abstract}
We derive an exact expression for the transmission amplitude of a
particle moving through a harmonically driven delta-function
potential by using the method of continued-fractions within the
framework of Floquet theory. We prove that the transmission
through this potential as a function of the incident energy
presents at most two real zeros, that its poles occur at energies
$n\hbar\omega+\varepsilon^*$ ($0<Re(\varepsilon^*)<\hbar\omega$),
and that the poles and zeros in the transmission amplitude come in
pairs with the distance between the zeros and the poles (and their
residue) decreasing with increasing energy of the incident
particle. We also show the existence of non-resonant "bands" in
the transmission amplitude as a function of the strength of the
potential and the driving frequency.

\end{abstract}

\maketitle
\section{Introduction}

Time-dependent potentials in mesoscopic systems have been studied
for a number of years in connection with electron-phonon
interactions ~\cite{Gelfand89}, quantum tunneling
time~\cite{Büttiker82,Stovneng89,Tanizawa96}, ionization
~\cite{Moiseyev91,Lebowitz00}, electronic transmission
~\cite{Bagwell92b,Cota93,Li99,Wagner94} and also in the field of
quantum chaos~\cite{Reichl92}. One of the interesting features of
localized time-periodic potentials is the presence of resonances
or quasi-bound "states", which could be thought of as electrons
dynamically trapped by the oscillating potential. This is also a
feature common to all multi-channel quantum scattering
problems~\cite{McVoy66,Bagwell92a}.

The solution of any three-term recursion relation, either in
functions or in operators, is a continued fraction (C.F.).  Such
is the case for tight-binding hamiltonians
~\cite{Kónya97,Davison97}, for the time-independent
Schr\"{o}dinger equation (in discretised space)
~\cite{Vigneron80}, and for harmonic time-dependent potentials,
such as an atom in a standing-wave laser field
~\cite{Autler55,Berg-Sorenson92} or for tunneling in the presence
of phonons ~\cite{Gelfand89}.

We first prove here, starting from Schr\"{o}dinger's equation,
that the transmission amplitude has the structure of a C.F. of
functions of the incident energy and the strength of the delta
potential. This is a considerable advantage over the numerical
computation of the transmission done before using this kind of
potential ~\cite{Bagwell92b,Cota93}. Our expression allows us to
study with greater detail, both analytically and numerically,
several different features of the transmission that had not been
noticed or explained before, such as the location of the zero-pole
resonances of the transmission and the almost periodic behavior of
their position, the existence of non-resonant "bands", the
dependence of the pole residues on energy and the existence of the
so-called (in the language of nuclear physics) "threshold
anomalies" in the transmission. We believe our work gives some
insight and clarifies issues in the general problem of scattering
through harmonically driven localized potentials. One such
potential is the Landau-B\"{u}ttiker potential, for which
numerical and analytical studies of the transmission have been
done ~\cite{Li99,Wagner94}, showing many similarities with the
transmission properties of the potential studied in this paper.

In Section 2 we use Floquet's theorem to derive the equations that
couple different components of the wave function in a plain wave
basis, and then we use these equations to find the S-matrix for
this potential. In Section 3 we solve the equations derived in
Section 2 and find the exact C.F. expression for the transmission
amplitudes. In Section 4 and 5 we study analytically and
numerically the zeros and poles of the transmission. In Section 6
we briefly discuss the "threshold anomalies" in the transmission
amplitude.
\section{Scattering Matrix and Transmission Amplitudes}

A formal treatment of the problem of scattering by a time-periodic
potential can be found in ~\cite{Li99,Saraga97}. Also, a
discussion about the sub-space of the Hilbert space suitable for
the treatment of time-periodic potentials can be found
in~\cite{Sambe72}.\newline The Hamiltonian we consider is
\begin{equation}\label{Hamiltonian}
H(x,t)=-\frac{\hbar^{2}}{2{\mu}}\frac{d^{2}}{dx^{2}}+V
\delta(x){\cos}(\omega t), \label{eq:ham}
\end{equation}
where ${\mu}$ is the mass of the particle. Even though energy is
not conserved, the Floquet energy, $\varepsilon$, is conserved for
this system and takes on a continuous range of values in the
interval $0{\leq}\varepsilon{\leq}{\hbar}{\omega}$. The Floquet
eigenstate with Floquet energy, $\varepsilon$, takes the form
\begin{equation}\label{psiexpansion}
{\Psi}_\varepsilon (x,t)={\sum_{n=-{\infty}}^{\infty}}~\psi_{n}(x)
e^{-\frac{i}{\hbar}(\varepsilon+n\hbar\omega)t}. \label{eq:3}
\end{equation}
Since the potential is zero everywhere except at $x=0$, we assume
$\psi_{n}(x)$ to be of the form
\begin{eqnarray}
{\psi_{n}}^{L}(x)=\frac{1}{\sqrt{k_{n}}}(a_{n}e^{i k_{n}x}+
d_{n}e^{-i k_{n}x})~~~{\rm for}~x<0 \nonumber\\
{\psi_{n}}^{R}(x)=\frac{1}{\sqrt{k_{n}}}(c_{n}e^{i k_{n}x}+
b_{n}e^{-i k_{n}x})~~~{\rm for}~x>0. \label{eq:5}
\end{eqnarray}
The factor $\frac{1}{\sqrt{k_{n}}}$ has been included to ensure
unitarity of the S-matrix and the wave-vectors $k_n$ are defined
by
\begin{eqnarray}
k_{n}=~~\sqrt{{2{\mu}\over {{\hbar}^2}}(\varepsilon+n\hbar\omega)}
\label{eq:7}
\end{eqnarray}
In this paper $n$ will always have the range $-\infty<n<\infty$.
The square root function has its branch cut on the real energy
axis (so that a real energy gives a real momentum) and for the
energy on the negative real axis we will use the Riemann sheet
that has $Im(k_{n})\geq0$, the so called "physical sheet". On this
sheet the momentum $k_{n}$ is on the positive imaginary axis for
any $n<0$. This is required to allow for evanescent modes
(exponentially decaying on both sides of the potential) since they
can contribute significantly to the wave-function in the
neighborhood of the delta function. These evanescent states are
also related to resonances or quasi-bound states which are known
to exist in multi-channel problems. To avoid any unphysical
exponentially growing states when dealing with positive pure
imaginary momentum states, we will require that $a_n=0$ and
$b_n=0$ for $n{\leq}-1$. To study the resonances in the
transmission we will allow the energy to take complex values and
will not necessarily stay on the physical sheet. This will be
discussed in section 5.

The Floquet eigenstate, ${\Psi}_\varepsilon (x,t)$, must be
continuous at $x=0$. This leads to the condition that
\begin{equation}
a_n+d_n=c_n+b_n. \label{eq:9}
\end{equation}
Because of the delta function in the Hamiltonian, the slope of
${\Psi}_\varepsilon (x,t)$ is discontinuous and satisfies the
condition
\begin{equation}
{d{\Psi}_\varepsilon \over
dx}{\biggr|}_{x=0^+}-{d{\Psi}_\varepsilon \over
dx}{\biggr|}_{x=0^-}={2{\mu}V\over
{\hbar}^2}{\cos}({\omega}t){\Psi}_\varepsilon (0,t) \label{eq:11}.
\end{equation}
This leads to the condition
\begin{equation}
c_n+d_n-b_n-a_n=-2i~[h_{n-1}(a_{n-1}+d_{n-1})+h_n(a_{n+1}+d_{n+1})],
\label{eq:13}
\end{equation}
where
\begin{equation}
h_n={{\mu}V\over 2{\hbar}^2\sqrt{k_nk_{n+1}}}. \label{eq:hn}
\end{equation}
We can now combine Eqs. (\ref{eq:9}) and (\ref{eq:13}), to obtain
the following relations between coefficients
\begin{equation}
c_n+ih_nc_{n+1}+ih_{n-1}c_{n-1}=a_n-ih_nb_{n+1}-ih_{n-1}b_{n-1}
\label{eq:15}
\end{equation}
and
\begin{equation}
d_n+ih_nd_{n+1}+ih_{n-1}d_{n-1}=b_n-ih_na_{n+1}-ih_{n-1}a_{n-1}.
\label{eq:17}
\end{equation}

It is useful to separate the propagating modes from the evanescent
modes. Let us define column vectors
\begin{equation}
{\bar c}_p=\pmatrix{c_0\cr c_1\cr c_2\cr \vdots\cr}~~~{\rm and}~~~
{\bar c}_e=\pmatrix{c_{-1}\cr c_{-2}\cr c_{-3}\cr \vdots\cr},
\label{eq:19}
\end{equation}
where ${\bar c}_p$ contains the amplitudes of the propagating
modes outgoing to the right, and ${\bar c}_e$ contains the
amplitudes of the evanescent modes that decay to the right.
Analogous definitions apply for the  column vectors ${\bar a}_p$,
${\bar b}_p$, ${\bar d}_p$, and ${\bar d}_e$.  Note that ${\bar
a}_e{\equiv}{\bar 0}$ and ${\bar b}_e{\equiv}{\bar 0}$. We can now
rewrite Eqs. (\ref{eq:15}) and (\ref{eq:17}) in the following form
\begin{eqnarray}
({\bar 1}_{pp}+{\bar X}_{pp}){\cdot}{\bar c}_p+{\bar
X}_{pe}{\cdot}{\bar c}_e= {\bar a}_p-{\bar X}_{pp}{\cdot}{\bar
b}_p, \nonumber\\ {\bar X}_{ep}{\cdot}{\bar c}_p+({\bar
1}_{ee}+{\bar X}_{ee}){\cdot}{\bar c}_e= -{\bar
X}_{ep}{\cdot}{\bar b}_p, \nonumber\\ ({\bar 1}_{pp}+{\bar
X}_{pp}){\cdot}{\bar d}_p+{\bar X}_{pe}{\cdot}{\bar d}_e= {\bar
b}_p-{\bar X}_{pp}{\cdot}{\bar a}_p, \nonumber\\ {\bar
X}_{ep}{\cdot}{\bar d}_p+({\bar 1}_{ee}+{\bar X}_{ee}){\cdot}{\bar
d}_e= -{\bar X}_{ep}{\cdot}{\bar a}_p, \label{eq:21}
\end{eqnarray}
where ${\bar 1}_{pp}$ and ${\bar 1}_{ee}$ are infinite dimensional
unit matrices and matrices ${\bar X}_{pp}$, ${\bar X}_{ee}$,
${\bar X}_{ep}$, and ${\bar X}_{pe}$ have matrix elements
\begin{eqnarray}
({\bar
X}_{pp})_{m,m'}=ih_m~({\delta}_{m,m'+1}+{\delta}_{m,m'-1}),~~~
({\bar X}_{ee})_{{\nu},{\nu}'}=ih_{\nu}~({\delta}_{{\nu},{\nu}'+1}
+{\delta}_{{\nu},{\nu}'-1}), \nonumber\\ ({\bar
X}_{pe})_{m,{\nu}}=ih_{-1}~{\delta}_{m,0}{\delta}_{{\nu},-1},\hspace{1.7cm}
({\bar
X}_{ep})_{{\nu},m}=ih_{-1}~{\delta}_{m,0}{\delta}_{{\nu},-1}.\hspace{1.0cm}
\label{eq:23}
\end{eqnarray}
Note that we have introduced the convention that the indices
$m=0,1,2,..,\infty$, and $\nu=-1,-2,-3..,-\infty$, to help
separate propagating modes from evanescent modes.

We can now write the scattering matrix for this system. The
scattering matrix, ${\bar S}$, connects the incoming propagating
modes to the outgoing propagating modes
\begin{equation}
\pmatrix{{\bar d}_p\cr {\bar c}_p\cr}={\bar
S}{\cdot}\pmatrix{{\bar a}_p\cr {\bar b}_p\cr}=\pmatrix{{\bar
r}&{\bar t'}\cr  {\bar t}&{\bar r'}\cr}{\cdot}\pmatrix{{\bar
a}_p\cr {\bar b}_p\cr}, \label{eq:25}
\end{equation}
where ${\bar t}$ and ${\bar t'}$ are matrices of transmission
probability amplitudes and ${\bar r}$ and ${\bar r'}$ are matrices
of reflection probability amplitudes.  More specifically, the
matrix element $t_{m,m'}=({\bar t})_{m,m'}$ is the probability
amplitude for the $m'$ mode entering from the left to transmit to
the right, and $r_{m,m'}=({\bar r})_{m,m'}$ is probability
amplitude that the $m'$  mode entering from the left and will
reflect to the left. Matrix elements of ${\bar r'}$ and ${\bar
t'}$ contain reflection and transmission coefficients for modes
entering  from the right.
 After some algebra one can show that
\begin{equation}
{\bar t}={\bar t'}=({\bar 1}_{pp}+{\bar Y}_{pp})^{-1} ~~~{\rm
and}~~~ {\bar r}={\bar r'}=-({\bar 1}_{pp}+{\bar
Y}_{pp})^{-1}{\cdot}{\bar Y}_{pp},
 \label{eq:26}
\end{equation}
where
\begin{equation}
{\bar Y}_{pp}={\bar X}_{pp}-{\bar X}_{pe}{\cdot}({\bar
1}_{ee}+{\bar Y}_{ee})^{-1}{\cdot}{\bar X}_{ep}.
 \label{eq:27}
\end{equation}
The effects of the evanescent modes are now explicitly included in
the scattering matrix.

For propagating modes entering in the $m^{th}$ channel on the
left, the total probability for transmission to the right is
\begin{equation}
T_m={\sum_{m'=0}^{\infty}}|t_{m',m}|^2.
 \label{eq:28}
\end{equation}
In the next section we focus on the transmission probability,
$T_0$. By expanding it in continued fractions, we can determine
all the transmission zeros and complex poles of this system.


\section{Continued fractions solution}

Let us now consider the special case of a single propagating mode,
entering in channel $m=0$ from the left and no propagating modes
entering from the right. In this case $a_n={\delta}_{n,0}$ and
$b_{n}=0$. The probability amplitude for the particle to be
transmitted into the $n^{th}$ channel (propagating or not) on the
right is $c_n$. For the propagating modes ($n\geq0$),
$c_{n}=t_{n,0}$. Let us now define the following quantity
\begin{equation}\label{f_n}
f_{n}=\frac{c_{n}}{c_{n+1}}~~~~~~~~{-\infty<n<\infty}
\end{equation}

In terms of $f_{n}$, Eq.(\ref{eq:15}) gives,

\begin{equation}\label{mastereq}
1+ \frac{i h_{n}}{f_{n}}+ i h_{n-1}f_{n-1}= 0~~~~~~{\rm
for}~{n{\neq}0}\\
\end{equation}
\begin{equation}\label{mastereq0}
1+ \frac{i h_{0}}{f_{0}}+ i h_{-1}f_{-1}=
\frac{1}{c_{0}}~~~~~~~{\rm when}~{n{=}0}\\
\end{equation}

For $n\geq0$ we can write the solution of (\ref{mastereq}) in the
form
\begin{equation}\label{nposi}
f_{n}=\frac{1}{-i h_{n}}(1+ \frac{i h_{n+1}}{f_{n+1}})
\end{equation}
or,
\begin{equation}\label{nposifraction}
f_{n}=\frac{1}{-i h_{n}}(\displaystyle
1+\frac{h_{n+1}^{2}}{\displaystyle
1+\frac{h_{n+2}^{2}}{\displaystyle 1+\frac{h_{n+3}^{2}}{\ddots}}})
\end{equation}

For $n\leq-1$ we write the solution in the form
\begin{equation}\label{nneg}
f_{n}=\frac{-i h_{n}}{1+i h_{n-1}f_{n-1}}
\end{equation}
or,
\begin{equation}\label{nnegfraction}
f_{n}=\frac{-i h_{n}}{\displaystyle
1+\frac{h_{n-1}^{2}}{\displaystyle
1+\frac{h_{n-2}^{2}}{\displaystyle 1+\frac{h_{n-3}^{2}}{\ddots}}}}
\end{equation}

\begin{figure}[t]
\fbox{\epsfig{file=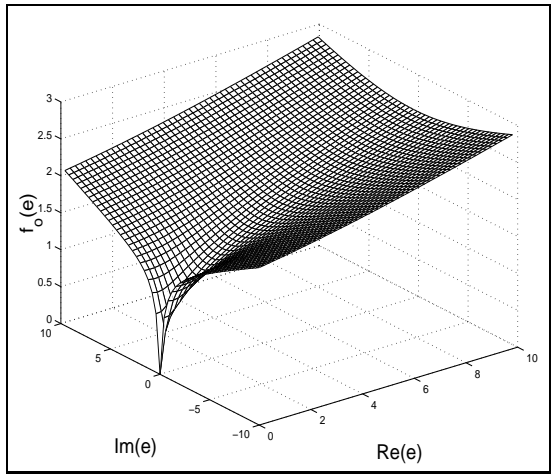,height=6cm,width=7cm}}
\caption{Graph of $|f_{0}(E)|^2$ in the complex plane (E in units
of $\hbar\omega$). $a=10$. Notice that
$f_{n}(\varepsilon)$=$f_{0}(n+\varepsilon)$=$f_{0}(E)$, therefore
this graph shows the behavior of $f_{n}(\varepsilon)$ for
$0\leq~n\leq 9$. The function behaves similarly, but with a
different scale, for other values of $a$.}\label{fig1}
\end{figure}

From the above expressions for the $f_{n}$'s (when $n=0$ and
$n=-1$) we can obtain $c_{0}$ from Eq.(\ref{mastereq0}),

\begin{equation}\label{c0}
c_{0}=t_{0,0}=\frac{1}{\displaystyle 1+ \frac{i
h_{0}}{\displaystyle f_{0}}+i h_{-1}f_{-1}}
=\frac{1}{\displaystyle 1+\frac{h_{0}^{2}}{\displaystyle
1+\frac{h_{1}^{2}}{\displaystyle 1+\frac{h_{2}^{2}}{\displaystyle
1+\frac{h_{3}^{2}}{\ddots}}}}+\frac{h_{-1}^{2}}{\displaystyle
1+\frac{h_{-2}^{2}}{\displaystyle
1+\frac{h_{-3}^{2}}{\displaystyle 1+\frac{h_{-4}^{2}}{\ddots}}}}}
\end{equation}
With the solution for the ${f_{n}}'s$ and $c_{0}$ known, any
coefficient $c_{n}$ can be found (using Eq.\ref{f_n}) in the
following way,
\begin{equation}\label{c_n's}
c_{n}=\frac{c_{0}}{f_{n-1}f_{n-2}...f_{0}}~~~{\rm for}~{n{\geq}1}
 $$and$$ c_{n}=f_{n}f_{n+1}....f_{-1}c_{0}~~~{\rm
for}~{n{\leq}-1}
\end{equation}
From here, the transmission probability, $T_0$, can be written as
\begin{equation}
T_0={\sum_{n=0}^{\infty}}|c_{n}|^{2}=
|t_{0,0}|^{2}(1+\sum_{n=1}{\frac{1}{ \prod_{n'=1}^{n}
|f_{n'-1}|^{2}}}) =|t_{0,0}|^{2}S(\varepsilon,V). \label{eq:31}
\end{equation}
As it can be seen in Fig.~\ref{fig1}, none of the $f_{n}'s$ (for
$n\geq0$) appear to have zeros in the half-plane
Re$[\varepsilon]>0$. This implies that the function
$S(\varepsilon,V)$ has neither zeros nor poles in that region;
consequently, the zeros and poles in the transmission probability
are the zeros and poles of $|t_{0,0}|^{2}$. From now on we
concentrate on $t_{0,0}$ only.
\section{Transmission Zeros}
The  coefficient $c_0$ can be taken to be a continuous function of
the incoming energy $E=\varepsilon+n\hbar\omega$, instead of a
function of the Floquet energy $\varepsilon$. This is so because,
when the incident energy is in channel $m$
($E=m\hbar\omega+\varepsilon$), we need to solve Eq.(\ref{eq:15})
with the condition $a_n=\delta_{n,m}$. The solution is given in
terms of the coefficient $c_m$ which now plays the former role of
$c_0$. The C.F. for $c_m$ is given by Eq.(\ref{c0}) with $m$ added
to all subscripts. As it can be seen easily,
$c_{m}(\varepsilon)=c_{0}(m\hbar\omega+\varepsilon)=c_{0}(E)$,
which means that the general solution of the problem, for any
incoming energy can be obtained from the C.F. expression for
$c_{0}$ derived in Eq.(\ref{c0}) evaluating it at any energy $E$.
Notice also that $c_0 (E)=
c_{0}(m\hbar\omega+\varepsilon)=t_{m,m}(\varepsilon)$. Because of
this, and to be rigorous with the notation, we define
$t_0(E)=t_0(m\hbar\omega+\varepsilon)\equiv
t_{m,m}(\varepsilon)=c_0(E)$. From this we can say that $t_0(E)$
contains the same information as the whole diagonal of the
transmission matrix $\overline{t}$ in Eq.(\ref{eq:25}).

To study the properties of $c_0(E)$ it is convenient to define the
following quantities (see Eq.~\ref{eq:hn})

\begin {equation}\label{gn}
g_{n}(\epsilon)\equiv{h_{n}}^{2}(\varepsilon)=\frac{a}{\sqrt{\epsilon+n}\sqrt{\epsilon+n+1}}
$$with dimensionless parameters$$
a\equiv\frac{mV^{2}}{8\hbar^{3}\omega},\hspace{2cm}
\epsilon\equiv\frac{\varepsilon}{\hbar\omega}$${\rm and} $$
e\equiv\frac{E}{\hbar\omega}=\epsilon+n \hspace{0.5cm} {\rm for}
\hspace{0.5cm} n\hbar\omega\leq E \leq (n+1)\hbar\omega
\end{equation}
Also, we define the function
\begin {equation}\label{F}
F_{0}(e)=F_{0}(n+\epsilon)\equiv F_{n}(\epsilon)\equiv 1+i
\frac{h_{0}(n+\epsilon)}{f_{0}(n+\epsilon)}
=1+\frac{g_{n}(\epsilon)}{\displaystyle
1+\frac{g_{n+1}(\epsilon)}{\displaystyle
1+\frac{g_{n+2}(\epsilon)}{\ddots}}}
\end{equation}
Notice that $g_n (\epsilon)$ also depends on $a$.

Using the definitions given above we can write the coefficient
$c_{0}(e)$ for the range  $n \leq e \leq (n+1)$, in the following
way

\begin{equation}\label{c0sec}
c_{0}(e)=c_{0}(n+\epsilon)=\frac{1}{\displaystyle F_n
(\epsilon)+\frac{g_{n-1}(\epsilon)}{\frac{\ddots}{\displaystyle
1+\frac{g_{0}(\epsilon)}{\displaystyle
1+\frac{g_{-1}(\epsilon)}{\displaystyle
1+\frac{g_{-2}(\epsilon)}{\ddots}}}}}}
\end{equation}
In the above expression all quantities are real (for $\epsilon$
real) except for $g_{-1}(\epsilon)$ which is pure imaginary. This
has an important consequence for the number of real zeros in
$c_{0}(e)$ as we shall see next.

Let us rewrite the equation above in a slightly different way:
\begin{equation}\label{proof2zeros}
c_{0}(n+\epsilon)=\frac{1}{\displaystyle
F_n(\epsilon)+\frac{g_{n-1}(\epsilon)}{\displaystyle
1+\frac{g_{n-2}(\epsilon)}{\frac{\ddots}{\displaystyle
1+\frac{g_{o}(\epsilon)}{\displaystyle 1+ i G(\epsilon)}}}}}
\end{equation}
where the continued fraction $G(\epsilon)$ is defined as
\begin{equation}
G(\epsilon)\equiv\frac{-a}{\displaystyle
\sqrt{1-\epsilon}\sqrt{\epsilon}(1+\frac{g_{-2}(\epsilon)}{\displaystyle
1+\frac{g_{-3}(\epsilon)}{\ddots}})}
\end{equation}
Notice $G(\epsilon)$ is a real function for $\epsilon$ real and
$0\leq\epsilon\leq1$.
\newline If Eq.~(\ref{proof2zeros}) is put in the form of a
standard fraction, we get
\begin{equation}
c_{0}(\epsilon)=\frac{1}{F_{0}+i G} ~~~~~~~~~~~~~~{\rm
for}~{n{=}0}$$and $$
 c_{0}(n+\epsilon)=\frac{i
P_{n}(g_{n-2},g_{n-3},...,g_{0})G+R_{n}(g_{n-2},g_{n-3}...,g_{0})}{i
Q_{n}(F_{n},g_{n-1},...,g_{0})G+S_{n}(F_{n},g_{n-1},...,g_{0})}~~~~~{\rm
for}~{n{>}0}
\end{equation}
In the last expression, $P_{n}, Q_{n}, R_{n}, S_{n}$ are
polynomials on the variables $g_{n}$ and $F_n$ indicated in
parentheses, such that the coefficient of every term is equal to
one and all variables appear elevated to the first power only
(i.e. $1+g_{1}+g_{1}g_{2}+..$). Notice that $G$ is the only
function that can take values between $-\infty$ to $\infty$. The
other functions ($F_n$ and $g_{n}$) are strictly positive and
finite for $\epsilon>0$ . This implies that $P_{n}, Q_{n},
R_{n},S_{n}$ are strictly positive and finite, which means that,
for $n>0$, the numerator of $c_0$ can never be zero and the
denominator does not go to infinity unless $G\rightarrow\infty$,
in which case the numerator would also blow up keeping the
fraction strictly positive. From this we can conclude that
\emph{there are no real transmission zeros for incident energies
$E>\hbar\omega$}.

\begin{figure}[t]
\fbox{\epsfig{file=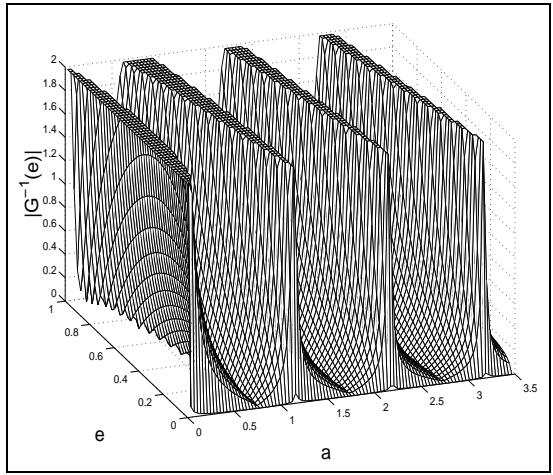,height=6cm,width=7cm}}
\caption{Graph of the function $|G^{-1}(\epsilon,a)|$. The zeros
of this function give the real zeros of the transmission.
}\label{fig3}
\end{figure}

\begin{figure}[t]
\epsfig{file=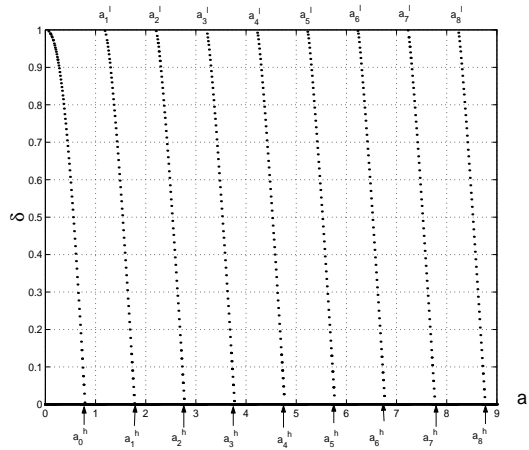,height=6cm,width=7cm} \caption{Location
of the real zero of $t_{0}$ as a function of $a$. \newline
$t_{0}(\delta,a)=G^{-1}(\delta,a)=0$.}\label{fig4}
\end{figure}

\begin{figure}[t]
\fbox{\epsfig{file=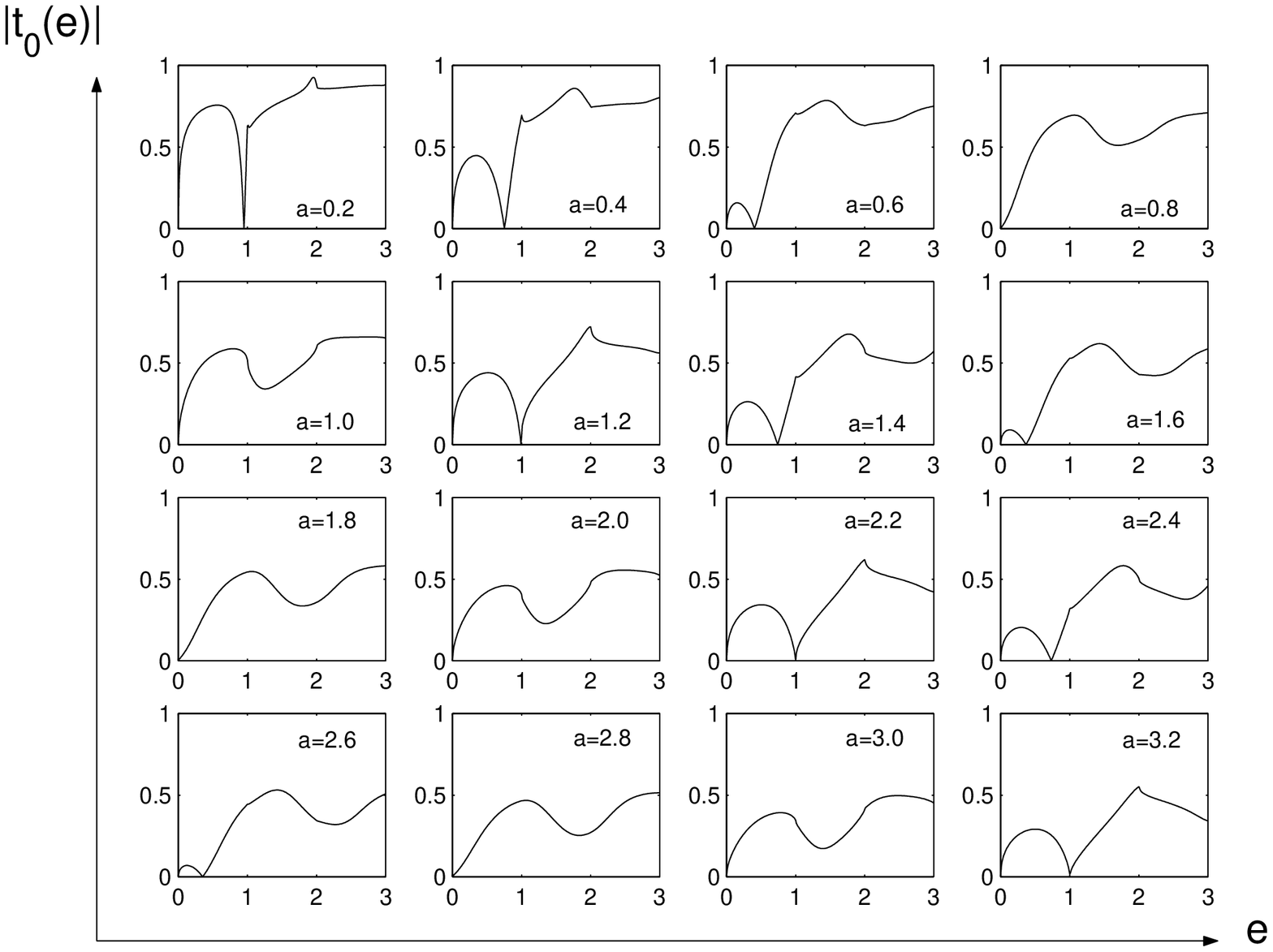,height=6cm,width=7cm}}
\caption{Sequence of $|t_{0}(e)|$ graphs for increasing values of
$a$ showing the evolution of the transmission real zeros.
}\label{fig5}
\end{figure}

For n=0 we see that a real zero can only happen when
$F_0(\epsilon,a)\longrightarrow \infty$ or
$G(\epsilon,a)\longrightarrow \infty$. The first case only occurs
when $\epsilon=0$ (G blows up at this point too). For the second
case, the real zeros of $t_{0}(\epsilon,a)$ are given by the zeros
of $G^{-1}(\epsilon,a)$. As it can be seen from Fig.~\ref{fig3}
the function $G^{-1}(\epsilon,a)$ seems to have some periodicity.
This can also be seen in Fig.~\ref{fig4} where the curves $\delta
(a)$ that satisfy $t_{0}(\delta,a)=G^{-1}(\delta,a)=0$ are shown.
This dependence of the transmission real zeros with the parameter
$a$ can be seen directly in Fig.~\ref{fig5} in the sequence of
transmission graphs for different values of the strength of the
delta.\newline It is interesting to notice in Fig.~\ref{fig4} and
Fig.~\ref{fig5} that there are intervals of $a$, around integer
values, for which the real zero in the transmission disappears. A
table (for $0\leq a \leq 9$) with the exact values (to seven
digits) of the intervals for which there is a real zero in the
transmission (apart from the trivial one at $E=0$) is shown in
Table~\ref{table1}. This behavior of the real zero seems to be
matched by the behavior of the poles of the transmission as we
will see next.
\section{Transmission Poles}
In this section we prove by induction that the poles of $c_{0}(e)$
occur at energies $e=(n+\epsilon^{*})$ with $0<Re(\epsilon^{*})<1$
and show that their residue decreases with increasing $n$
(energy). We start with equation (\ref{c0sec}) and write it in the
form
\begin{equation}\label{c0FG}
c_{0}(n+\epsilon)=\frac{1}{F_{n}(\epsilon)+G_{n}(\epsilon)}
\end{equation}
where $F_{n}$ and $G_{n}$ satisfy
\begin{equation}\label{FG}
F_{n}(\epsilon)=1+\frac{g_{n}(\epsilon)}{F_{n+1}(\epsilon)}$$ $$
G_{n}(\epsilon)=\frac{g_{n-1}(\epsilon)}{1+G_{n-1}(\epsilon)}
\end{equation}
Let's assume that $c_{0}(e)$ has a pole at $e=n+\epsilon^{*}$.
This implies that $c_{0}(e)$ has also a pole at
$e=n+1+\epsilon^{*}$ because we can write from Eqs.~(\ref{c0FG})
and~(\ref{FG})
\begin{equation}\label{c0poles}
c_{0}(n+1+\epsilon) =\frac{1}{F_{n+1}(\epsilon)+G_{n+1}(\epsilon)}
$$ $$ ~~~~~~~~~~~~~~~~~~~~~~~=\frac{1}{\displaystyle
\frac{g_{n}(\varepsilon)}{F_{n}(\epsilon)-1}+\frac{g_{n}(\epsilon)}{1+G_{n}(\epsilon)}}$$
$$~~~~~~~~~~~~~~~~~~~~~~
=\frac{(1+G_{n}(\epsilon))(F_{n}(\epsilon)-1)}{g_{n}(\epsilon)(F_{n}(\epsilon)+G_{n}(\epsilon))}$$
$$
~~~~~~~~~~~~~~~~~~~~~~~~=\frac{(1+G_{n}(\epsilon))}{F_{n+1}(\epsilon)(F_{n}(\epsilon)+G_{n}(\epsilon))}
\end{equation}
It is important to notice that, since $f_0(e)$ (as shown in
Fig.\ref{fig1}) has neither poles nor zeros in the half plane
$Re(e)>0$, the function
$F_n(\epsilon)=F_0(n+\epsilon)=1+i\frac{h_0(e)}{f_0(e)}$
 ~does not have poles and $F_n(\epsilon)\neq 1$. This
implies that (from the last line of Eq.~(\ref{c0poles})) if
$c_{0}(e)$ has a pole at $n+\epsilon^{*}$ (which means that
$F_n(\epsilon^{*})+G_n(\epsilon^{*})=0$), then it must have a pole
at $n+1+\epsilon^{*}$ (The numerator in Eq.~(\ref{c0poles}) does
not vanish at $\epsilon=\epsilon^{*}$ because
$G_n(\epsilon^{*})=-F_n(\epsilon^{*})\neq -1$).

Notice also that when $n\longrightarrow\infty$,
$F_{n}\longrightarrow 1$, because
$f_0(n+\epsilon)\rightarrow\infty$, and
~$h_0(n+\epsilon)\rightarrow 0$ in this limit. This means that the
location of the transmission zero approaches the location of the
pole as the incoming energy grows; the zero happens when
$\epsilon=\delta$, with $G_{n}(\delta)=-1$; the pole occurs when
$\epsilon=\epsilon^*$, with $
G_{n}(\epsilon^*)=-F_{n}(\epsilon^*)$. Obviously
$\delta\rightarrow\epsilon^{*}$ because
$F_{n}(\epsilon^*)\longrightarrow 1$ as $n\rightarrow\infty$. From
that we conclude that the residue of the poles in the transmission
amplitude tend to zero as $n\longrightarrow\infty$. This explains
why, even though the transmission amplitude has an infinite number
of poles separated by a distance of $\hbar\omega$ in the incoming
energy, all at the same distance from the real axis, only the
effect of the first poles can be seen in the graphs of
transmission probability versus incoming energy. Another way to
say this is that the zeros (in the complex plane) at higher
energies are very close to the poles, therefore canceling out the
possible effect of the poles in the transmission.

\begin{figure}[t]
\fbox{\epsfig{file=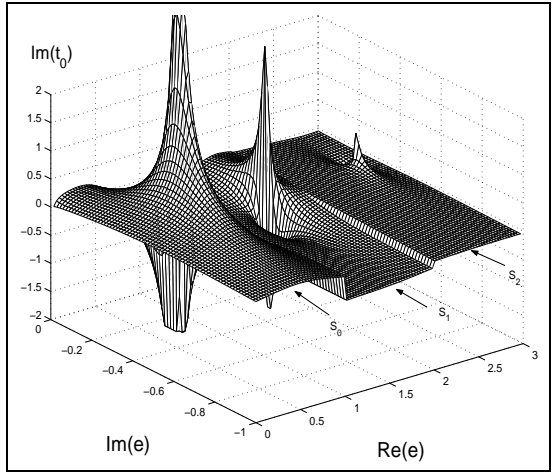,height=6cm,width=7cm}} \caption{
Graph of the imaginary part of $t_{0}(e)$ for a=0.5,
 showing several poles. Each pole is located on a different sheet.
}\label{fig6}
\end{figure}

\begin{figure}[t]
\fbox{\epsfig{file=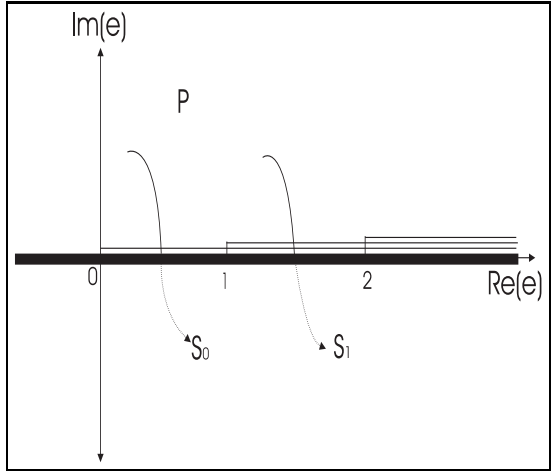,height=6cm,width=7cm}} \caption{
top view of the physical sheet P and the threshold branch points
and cuts in the complex energy plane. Indicated with a thin line
are the branch cuts corresponding to the branch points $n=0,1,2$;
the thick line represents all the branch cuts with branch point at
negative energy($n<0$ branch points). Paths that start on the
upper half of P and go under the branch cut lead to the different
sheets $S_{n}$ (which in this figure are assumed to be underneath
P).}\label{fig7}
\end{figure}

\begin{figure}[t]
\fbox{\epsfig{file=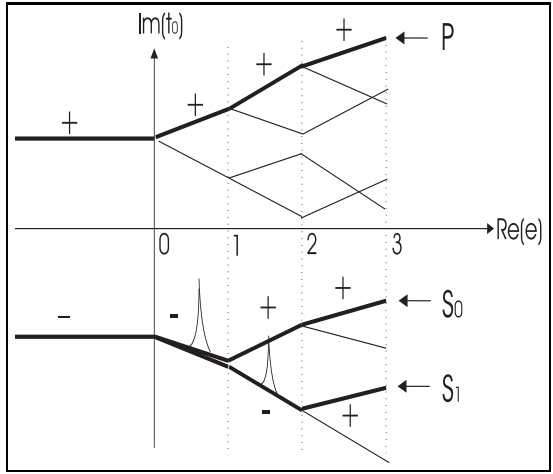,height=6cm,width=7cm}}
\caption{Sketch of $Im(t_{0}(e))$ versus $Re(e)$ (we assume
$Im(e)=constant<0$) which gives a schematic representation of the
different sheets where the poles of the transmission are located.
The peak structures in the lower part of the figure represent
poles. }\label{fig8}
\end{figure}

In Fig.~\ref{fig6} ~we show a graph of the imaginary part of
$t_{0}(e)$ where the poles in channels 0, 1, 2, can be seen (we
refer to channel $n$ as the strip in the complex energy plane that
satisfies $n<Re(e)<n+1$). It is evident in that graph that the
poles have support on different sheets; we will call these sheets
$S_n$, where $n$ refers to the channel number. This sheeted
structure comes from the fact that, because the functions
$k_{n}(e)$ have a branch point at $e=-n$ and two Riemann sheets,
any function of $k_{n}(e)$ ($t_{0}(e)$ in particular) will have a
multiple-sheeted structure. What is called the "physical" sheet
(P) in the context of multiple channel scattering is the sheet
obtained when selecting all the Riemann sheets with Im$(k_{n})>0$.
Each sheet of the full multi-sheeted surface can be labeled by the
sequence of signs of Im($k_{\infty}$),..., Im($k_{1}$),
Im($k_{0}$), Im($k_{-1}$),..., Im($k_{-\infty}$)., e.g., $(+..-
$\textbf{+}$+..+)$~ (bold has been used to indicate the sign of
Im($k_{0}$)); fortunately we do not need to consider all these
sheets, since only a small fraction of them are of physical
interest; the principal one being P or (+..+$\textbf{+}$+..+).
Crossing the real axis from P at an energy $m<e<m+1$ crosses all
the branch cuts whose branch point is at an energy smaller than
$e=m$ (see figure ~\ref{fig7}) . The sheet that is the smooth
continuation of P into the lower complex plane (smooth even at the
cut) is the one obtained by taking Im($k_{-n}$)$<0$ for all $n\leq
m$ and Im($k_{-n}$)$>0$ for all $n> m$ . We call these sheets
$S_m$ and they are precisely the ones where the poles in the
transmission are found, as shown schematically in Fig.~\ref{fig8}.

Resonances, as opposed to bound state poles, do not occur in the
"physical" sheet; however, from the unitarity of the S-matrix
extended to the complex plane, they can have important effects on
the behavior of the S-matrix along the real axis (see
\cite{McVoy66} for an excellent discussion on multiple channel
scattering).

In single channel scattering, the unitarity of the S-matrix on the
real axis has consequences in the analytic structure of the
S-matrix in both sheets, the most important one being the
zero-pole structure of the resonances: every pole has a companion
zero at a position complex conjugate to the position of the pole
but on a different sheet. For multiple channel scattering the
position of the pole and zero are not so simply related and often
the pole and zero appear in the same sheet.

\begin{figure}[t]
\fbox{\epsfig{file=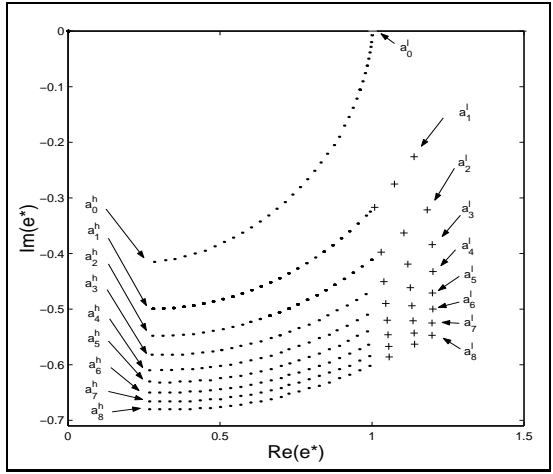,height=6cm,width=7cm}}
\caption{Trajectory of one of the transmission poles as $a$ is
changed from 0 to 9. Here $a_{n}^{l}/a_{n}^{h}$ refers to the
lower/higher value of the interval $a_{n}$ given in Table 1.
}\label{fig9}
\end{figure}

\begin{figure}[t]
\fbox{\epsfig{file=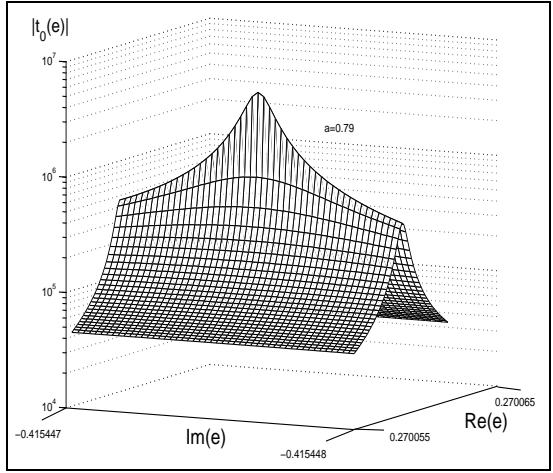,height=6cm,width=7cm}}
\caption{false "pole" of $t_{0}(e)$ for $a=0.79$, right after the
zero has disappeared at $\delta=0$ for $a=a_0^h$.}\label{fig10}
\end{figure}

\begin{figure}[t]
\fbox{\epsfig{file=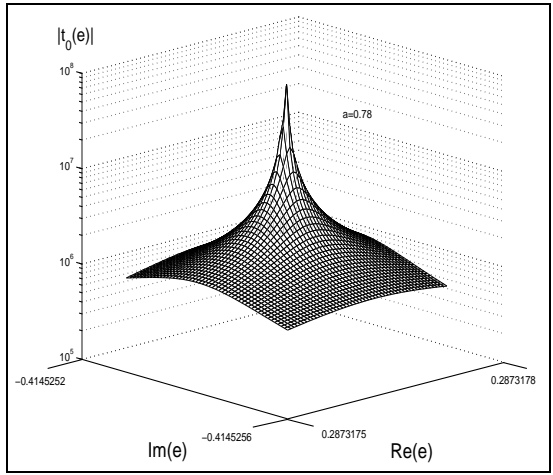,height=6cm,width=7cm}}
\caption{True pole of the transmission for $a=0.78$, slightly
lower than the value of a ($a=a_0^h$) for which the zero
disappears.}\label{fig11}
\end{figure}

The position of the pole ($e^{*}$) in the first channel, can be
found by looking at the zeros of the function
$F_{0}(e,a)+G_{0}(e,a)$ with the appropriate selection of Riemann
sheets, as described before, so as to be on the sheet $S_0$. A
graph of the position of the pole as the strength of the potential
is varied is shown in Fig.~\ref{fig9}, where it can be seen that
the pole approaches the real axis at $e^* =1$ when
$a\rightarrow0$, annihilating the real zero which also goes to the
same position with $a\rightarrow0$. As the parameter $a$ is
increased the zero moves to the left on the real axis (see
Fig.~\ref{fig4}) and the pole moves away from the real axis
describing an arch. At $a=a_0^h=0.7821147$ (all values of a are
accurate to the seventh digit) the zero disappears at $\delta=0$
and so does the pole. For values of $a$ slightly greater, the
transmission has a very high peak but it is not a pole, as can be
seen in Fig.~\ref{fig10}~(compare with the pole at $a=0.78$ in
Fig.~\ref{fig11}). In the interval $a_0^h<a<a_1^l$ the
transmission does not have any zero-pole resonances. In this
interval the peak or false "pole" follows a trajectory that seems
the continuation of the trajectory the pole had followed until it
vanished. When $a=a_1^l$ the zero reappears on this channel,
entering at $\delta=1$. At this value however, we do not see any
pole in this channel. There is in fact a pole on $S_0$ but it is
located in the next channel ($1< Re(e^*)<2$). For $a\approx 1.3$
the pole has finally made it into the $n=0$ channel. As $a$ is
further increased the pole continues to describe an arch until it
disappears along with the zero. A similar behavior to the one
described above occurs for increasing values of the parameter $a$,
with the pole describing ever increasing arches, each one farther
away from the real axis than the previous one. An identical
behavior of the poles on other sheets is expected from the
discussion following Eq.(\ref{c0poles}).

This quasi-periodic behavior of the transmission resonances as a
function of the strength of the oscillating potential and the
driving frequency has not been found before. Around integer values
of $a$ the oscillating potential seems to be incapable of
dynamically trapping particles not even for a short period of
time.

\section{Threshold anomalies}
\begin{figure}[h]
\fbox{\epsfig{file=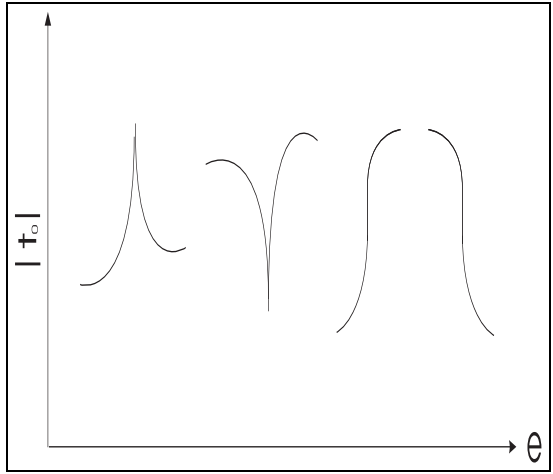,height=6cm,width=7cm}}
\caption{Various forms of transmission amplitude behavior at a
channel opening. }\label{fig12}
\end{figure}

\begin{figure}[h]
\fbox{\epsfig{file=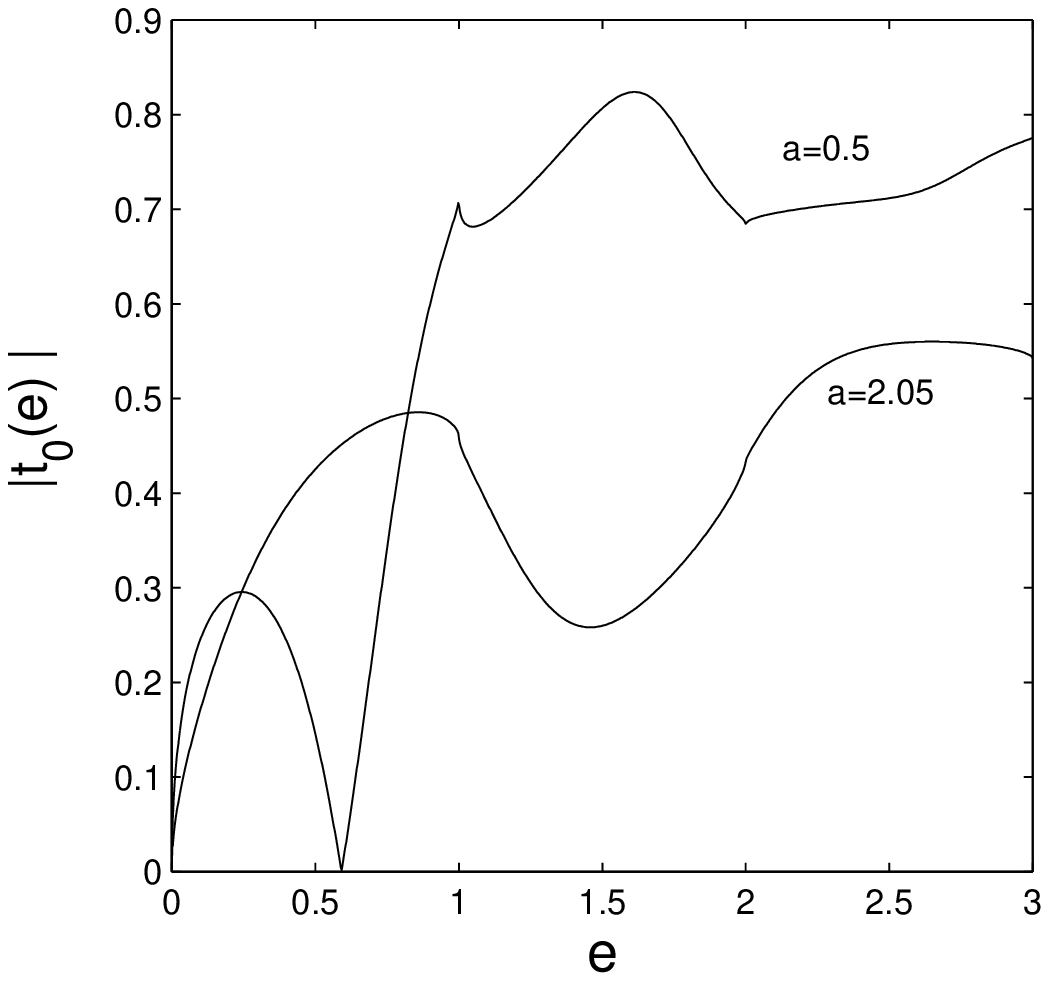,height=6cm,width=7cm}}
\caption{Graph of $|t_{0}(e)|$ vs. incoming energy. Threshold
anomalies of "cusp" type at the channel openings for a=0.5 .
Threshold anomalies of "rounded step" type for a=2.05
.}\label{fig2}
\end{figure}

One of the most evident features in the transmission versus energy
curves for all values of $a$ is what seems to be a discontinuity
in the slope at $e=n\hbar\omega$ (channel openings). A close look
into this regions reveals that there is a rapid divergence of the
slope at the thresholds, that is,
$\frac{dt_{0}}{de}\rightarrow\pm\infty$ as
$e\rightarrow~n\hbar\omega$. This is actually a common occurrence
in all multi-channel scattering problems, and it can be traced
back to the fact that when the energy is in the $n^{th}$ channel
near threshold, the first open-channel momentum is
$k_{-n}=\sqrt{e-n\hbar\omega}$.  It can be proven that the
S-matrix has elements that are linear functions of this momentum
near threshold. This clearly implies that the derivative of these
elements with respect to the energy must diverge at threshold (for
more details see ~\cite{Newton66}). These threshold anomalies can
easily be  proven to exist in our particular time dependent
potential by looking at our C.F. solution for $t_{0}(e)$ in
Eq.~\ref{c0}. From it, it is clear that the derivative of this
expression gives an infinite number of terms, each one
proportional to the derivative of some function $g_{n}(e)$. These
functions and their derivatives diverge at their thresholds (see
the expression for $g_n$ in Eq.~\ref{gn}), therefore the
derivative of $t_{0}(e)$ must also diverge at each threshold.
According to this, the shape of the threshold anomaly can be of
four different kinds, two cusps like and two rounded steps, as
shown in Fig.~\ref{fig12}. These four different kinds of anomalies
are shown in Fig.~\ref{fig2} as they occur in the $|t_{0}(e)|$
graphs for two different values of $a$.
\section{Conclusions}
The exact solution for the problem of transmission of a particle
through a monochromatic oscillating potential has been found with
the use of C.F's. This approach represents a significant
improvement over the standard method which can only be carried out
numerically since it involves the inversion of an infinite
tri-diagonal matrix, which can be a computationally demanding
problem when studying the zero-pole structure of the transmission
on the complex plane. The efficiency of the C.F. method allowed us
to perform, in a short time, calculations that required including
processes of phonon emission and absorption of 65th order(for
$a=9$) which would have required in the standard method the
computation of a 131x131 S-matrix for every point in the complex
plane where the transmission is to be evaluated. Most importantly,
it allowed us to prove rigorously some general properties of the
transmission for this system, such as the existence of
transmission zeros only in the first channel, the location of the
poles at regular intervals of $\hbar\omega$ in the incident
energy, the decrease in the residue of the poles with increasing
energy, and also allowed us to understand the existence of
threshold anomalies at the channel openings.

The study of the behavior of the poles and real zeros as a
function of the parameter $ a=\frac{mV^{2}}{8\hbar^{3}\omega}$
showed an interesting quasi-periodic dependence on $a$ and the
presence of "bands" of non-resonant values of $a$ for which the
transmission does not have zero-pole resonances. It can also be
mentioned that the life-time of the quasi-bound states decreases
with increasing $a$, which can be accomplished by either
increasing the strength $V$ of the potential or by decreasing the
frequency of the oscillations.\newline

The scattering of particles through a time-periodic potential is
another example of multi-channel scattering, for which a great
deal of theory and research has been done in the past and still
continues to be an important topic of research. In this paper we
have put Floquet scattering in the greater context of
multi-channel scattering to which it belongs and by doing so we
believe some aspects of the transmission through oscillating
(localized) potentials have been clarified.

Even though these results where derived for an oscillating delta
potential, a similar behavior can be expected to be found for
other oscillating finite-range potentials such as the
Landauer-B\"{u}ttiker potential, since this potential reduces to
the oscillating delta in an appropriate limit, and also because
several common features can already be observed when comparing the
transmission curves for this two potentials (see~\cite{Li99}) .
The exact solution for the oscillating delta plus a static
(attractive or repulsive) delta potential can be found using the
same C.F. method. The transmission through this potential has been
studied numerically before~\cite{Bagwell92b}, showing features
similar to the ones discussed here with the main difference been
that the location of the resonances is displaced by an amount of
energy roughly equal to the energy of the bound state associated
with the static part of the potential.\newline \clearpage

\section{Acknowledgements}

The authors wish to thank the Welch Foundation, Grant No.F-1051,
NSF Grant INT-9602971  and DOE contract No.DE-FG03-94ER14405 for
partial support of this work.
\bigskip

\maketitle

\clearpage
\begin{table}[t]
\begin{tabular}{c|c|c| }
$ $ & $l$ & $h$ \\
 \hline $a_{0}$ & 0         & 0.7821147\\
 \hline $a_{1}$ & 1.1652568 & 1.7710590\\
 \hline $a_{2}$ & 2.1873508 & 2.7667368\\
 \hline $a_{3}$ & 3.1979937 & 3.7642963\\
 \hline $a_{4}$ & 4.2045658 & 4.7626808\\
 \hline $a_{5}$ & 5.2091415 & 5.7615115\\
 \hline $a_{6}$ & 6.2125623 & 6.7606150\\
 \hline $a_{7}$ & 7.2152455 & 7.7598995\\
 \hline $a_{8}$ & 8.2174231 & 8.7593115\\
 \hline
\end{tabular}
\caption{Intervals of a=$\frac{mV^{2}}{8\hbar^{3}\omega}$ for
which a zero-pole resonance can be found in the system. The
notation $a_{n}^{l}/a_{n}^{h}$ (used in the text and in some
graphs) refers to the lower/higher value of the interval
$a_{n}$.}\label{table1}
\end{table}
\clearpage





\begin{thebibliography}{aa}
\bibitem{Gelfand89}B.Y. Gelfand, S. Schmitt-Rink, A.F.J. Levi, Phys.Rev.Lett., 62, 1683(1989).
\bibitem{Büttiker82}M. B\"{u}ttiker, R. Landauer, Phys.Rev.Lett., 49, 1739(1982).
\bibitem{Stovneng89}J.A. Stovneng, E.H. Hauge, J.Stat.Phys., 57, 841(1989).
\bibitem{Tanizawa96}T. Tanizawa, J.Phys.Soc.Japan, 65, 3157(1996).
\bibitem{Moiseyev91}F. Bench,H.J. Korsch, N. Moiseyev, J.Phys.B,
24, 1321(1991).
\bibitem{Lebowitz00}O. Costin, J.L. Lebowitz, A. Rokhlenko, J.Phys.A, 33, 6311(2000).
\bibitem{Bagwell92b}P.F. Bagwell, R.K. Lake, Phys.Rev.B, 46, 15329(1992).
\bibitem{Cota93}E. Cota, J.V. Jose, F.Rojas,
Nanostructured Materials, 3, 349(1993).
\bibitem{Li99}W. Li, L.E. Reichl, Phys.Rev.B, 60, 15732(1999).
\bibitem{Wagner94}M. Wagner, Phys.Rev.A, 51, 798(1995).
\bibitem{Reichl92}L.E. Reichl in {\it{The Transition to Chaos in
Conservative Classical Systems: Quantum Manifestations}},
(Springer-Verlag, New York, 1992).
\bibitem{McVoy66}K.W. McVoy in {\it{Fundamentals in Nuclear
Theory}}, edited by A. De-Shalit and C. Villi, (International
Atomic Agency, Vienna, 1967).
\bibitem{Bagwell92a}P.F. Bagwell, A. Kumar,
R.K. Lake in {\it{Quantum Effect Physics,
 Electronics and applications}}, edited by K. Ismail et
al., (Adam-Hilger, Bristol, 1992).
\bibitem{Kónya97}B. K\'{o}nya, G. L\'{e}vai, Z. Papp, J.Math.Phys., 38, 4832(1997).
\bibitem{Davison97}S.G. Davison, R.A. English, Z.L. Miskovi\v{c}, F.O. Goodman, A.T. Amos,
and B.L. Burrows, J.Phys.-Cond.Mat. 9, 6371(1997).
\bibitem{Vigneron80}J.P. Vigneron, Ph. Lambin, J. Phys.A: Math. Gen.
13, 1135(1980).
\bibitem{Autler55}S.H. Autler, C.H. Townes, Phys.Rev., 100, 703(1955).
\bibitem{Berg-Sorenson92}K. Berg-Sorenson, Y. Castin, E. Bonderup, K. Molmer, J.Phys.B, 25, 4195(1992).
\bibitem{Saraga97}D.S. Saraga, M. Sassoli-de-Bianchi, Helv.Phys.Acta, 70, 751(1997).
\bibitem{Sambe72}H. Sambe, Phys.Rev.A, 7, 2203(1972).
\bibitem{Newton66}R.G. Newton in {\it{Scattering theory of waves and
particles}}, (McGraw-Hill, New York, 1966).
\end{thebibliography}
\end{document}